\documentclass{aa}
\usepackage{graphicx}
\usepackage{longtable}

\usepackage{txfonts}
\begin{document}

\title{WINGS-SPE\\ Spectroscopy in the WIde-field Nearby Galaxy-cluster Survey.}
\titlerunning{Spectroscopy in WINGS}

   \author{A. Cava \inst{1,2,3}\thanks{E-mail: acava@iac.es}, D. Bettoni  \inst{1}
          \and B.~M. Poggianti\inst{1} \and W.~J. Couch\inst{4}  \and M. Moles\inst{5}\and J. Varela\inst{1} \and A. Biviano\inst{6} \and M. D'Onofrio\inst{7} \and A. Dressler\inst{8} \and G. Fasano\inst{1} \and J. Fritz\inst{1} \and P. Kj\ae rgaard\inst{9} , M. Ramella\inst {6} and T. Valentinuzzi\inst{7} 
          }

   \institute{INAF\,-\,Osservatorio Astronomico di Padova, Vicolo dell'Osservatorio 5, 3512, Padova, Italy
    \and
    Universit\`a degli Studi di Milano, Dipartimento di Fisica, Via Celoria 16, 20133 Milano, Italy
    \and
    Instituto de Astrofísica de Canarias, C/Vía Lactea s/n, 38200 La Laguna, Tenerife, Spain
    \and
    Centre for Astrophysics and Supercomputing, Swinburne University of Technology, Hawthorn VIC 3122, Australia
    \and
 Instituto de Astrof\'\i sica de Andaluc\'\i a (C.S.I.C.) Apartado 3004, 18080 Granada, Spain       
 \and INAF\,-\, Osservatorio Astronomico di Treste ,Via Tiepolo 11, 34143 Trieste, Italy 
  \and Universit\`a degli Studi di Padova, Dipartimento di Astronomia, Vicolo Osservatorio 3, 35122 Padova, Italy     
 \and
 Observatories of the Carnegie Institution of Washington, Pasadena, CA 91101, USA
\and
 Copenhagen University Observatory. The Niels Bohr Insitute for Astronomy Physics and Geophysics, Juliane Maries Vej 30, 2100 Copenhagen, Denmark }

   \date{Received; accepted}
  \abstract
   {}
   {We present the results from a comprehensive spectroscopic survey of the WINGS (WIde-field Nearby Galaxy-cluster Survey) clusters, a program called WINGS-SPE. The WINGS-SPE sample consists of 48 clusters, $22$ of which are in the southern sky and $26$ in the north. The main goals of this spectroscopic survey are: (1) to study the dynamics and kinematics of the WINGS clusters and their constituent galaxies, (2) to explore the link between the spectral properties and the morphological evolution in different density environments and across a wide range in cluster X-ray luminosities and optical properties.}
   {Using multi-object fiber-fed spectrographs, we observed our sample of WINGS cluster galaxies at an intermediate resolution of 6-9 \AA\ and, using a cross-correlation technique, we measured redshifts with a mean accuracy of $\sim45$\,km\,s$^{-1}$.}
   {We present redshift measurements for $6,137$ galaxies and their first analyses. Details of the spectroscopic observations are reported. The WINGS-SPE has $\sim 30 \%$ overlap with previously published data sets, allowing us to do both a complete comparison with the literature and to extend the catalogs.}
   {Using our redshifts, we calculate the velocity dispersion for all the clusters in the WINGS-SPE sample. We almost triplicate the number of member galaxies known in each cluster with respect to previous works. We also investigate the X-ray  luminosity vs. velocity dispersion relation for our WINGS-SPE clusters, and find it to be consistent with the form $L_x \propto \sigma_v^4$.}

   \keywords{Galaxies: clusters: general -- Galaxies: distances and redshifts
               }

   \maketitle

\section{Introduction}

Rich clusters of galaxies have long been recognized as  valuable tools with which to study cosmology and galaxy 
formation. They are the most massive virialized objects in the universe and, as such, provide excellent laboratories for studying the influence of environment on galaxy formation and evolution, as manifested by the morphology-density relation (\cite{Dressler80}, \cite{Dressler97}, \cite{Fasano00}, \cite{Postman05}). The identification of the properties and content of clusters of galaxies is able to clarify their cosmic evolution since they can be detected at large distances. However clusters of galaxies are now recognized to not be simple relaxed structures; rather, they are evolving via merging processes in a hierarchical fashion from poor groups to rich clusters. In fact in the cluster outskirts, galaxies are still falling in for the first time, so it is possible also to explore environmental effects over a wide dynamic range in density. 

Clusters contain large populations of galaxies at a common distance, which can be used to derive redshift-independent relative distance estimates (\cite{Dressler87}). This allows the study of deviations from pure Hubble flow, i.e., the peculiar velocity field, and hence the dark matter distribution in the local universe (e.g., \cite{LB88}; see also \cite{dekel}; \cite{strauss} for reviews).  Cluster velocity dispersions provide a measure of cluster mass (\cite{fisher98}; \cite{tran99}; \cite{bor99a}; \cite{lubin02}). The measurement of cluster velocity dispersions should be made using statistics insensitive to galaxy redshift outliers and the shape of the velocity distribution. However, the uncertainties in these studies are large, and there remains the possibility of systematic errors due to the heterogeneity of the spectroscopic data sets available for nearby clusters ($z<0.1$). In recent years,  a systematic investigation of a large sample of nearby clusters has become possible due to the advent of CCD mosaics and high multiplex multi-object fiber-fed spectrographs, allowing photometric and spectroscopic observations over large solid angles (\cite{noaoI}, \cite{fasano06}).

In the last few years, large redshift surveys such as the 2dF Galaxy Redshift Survey (hereafter 2dFGRS; \cite{DePropris}) and especially the Sloan Digital Sky Survey (SDSS, \cite{Goto}; \cite{Bahcall}; \cite{Miller}) have been the primary source for the compilation of data on nearby clusters of galaxies. However, both these surveys have their limitations. The 2dFGRS is only a spectroscopic survey while the SDSS, being a large area survey, is not deep enough to study the fainter end of the luminosity function. More recently, the NOAO Fundamental Plane Survey (\cite{noaoI}) started to address these shortcomings, although its primary goal is not to study the evolution of galaxies, but rather to trace large scale velocity fields using the Fundamental Plane.

The need for accurate redshift measurements has also become evident from the great progress that has been made in recent years in the observations of the signatures of cluster merging processes. The presence of substructure, which is indicative of a cluster in an early phase of the process of dynamical relaxation or of secondary infall of clumps into already virialized clusters, occurs in a large percentage of clusters (\cite{drshec}; \cite{west}; \cite{rhee}; \cite{bird}; \cite{escalera}; \cite{west1}; \cite{girardi97}; \cite{solanes}; \cite{biviano}; \cite{burgett}; \cite{flin}). Studying cluster substructure therefore allow us to investigate the process by which clusters form. In addition, it also enable us to better understand the mechanisms affecting galaxy evolution in clusters, which can be accelerated by the effects of a cluster-subcluster collision. An interesting result has been produced by our recent study (\cite{ramella07})  in which we point out that the fraction of clusters with sub-clusters in the WINGS sample (73\%) is higher than in most previous studies. In addition, in a following paper of the WINGS-SPE series (Cava et al. 2009, in prep.) a more detailed analysis of the sub-clustering in the WINGS sample will be included, exploiting and comparing different 2-D and 3-D approaches.

In this paper we present redshifts for galaxies in 48 clusters belonging to the Wide-field Imaging Nearby Galaxy-cluster Survey (WINGS) sample. In addition, we present cluster velocity dispersion measurements derived from our redshift data. These clusters comprise an almost complete X-ray selected sample of galaxy clusters at $z=0.04-0.07$ (see \cite{fasano06} for details of the survey).  

The main goal of the WINGS-SPE spectroscopic follow-up program is to supply a complete and uniform set of spectroscopic data such as redshift (present paper), line indices, and line widths useful to investigate the dynamics of the clusters and derive star formation histories, star formation rates (Fritz et al., \cite{fritz}, and Fritz et al. in prep.), and other structural and physical properties for cluster galaxies. These new data will shed more light on the link between the evolution of star formation and galaxy morphology, as well as the dependence on the characteristics of the cluster and where the galaxies are located within the cluster. Furthermore, the availability of data on low redshift clusters can be used as a present-day reference for studies of distant clusters (see e.g. Poggianti et al. \cite{poggianti06}). 

This paper is organized as follows: In \S 2 we give general information on the WINGS-SPE objectives and working strategy; in \S 3 we present the spectroscopic observations and describe the reduction processes; in \S 4 redshifts measurements and catalogs are presented, while in \S 5 we check the data quality performing a comparison with data in the literature. Finally we draw our conclusions in \S 6.

\section{Survey strategy}
\label{sec2}

WINGS clusters (see \cite{fasano06} for details) have been selected from three X-ray flux limited samples compiled from ROSAT All-Sky Survey data: the ROSAT Brightest Cluster Sample (\cite{ebeling98}, BCS), and its extension (\cite{ebeling00}, eBCS) in the northern hemisphere and the X-Ray-Brightest Abell-type Cluster sample (\cite{ebeling96}, XBACs) in the southern hemisphere. 

The global sample contains 77 clusters (36 in the northern hemisphere and 41 in the south) over a broad range of richness, Bautz-Morgan class and X-ray luminosities. Our aim was to obtain spectroscopic data for the whole sample. However, bad weather conditions (we lost $\sim 25\%$ of our observing time, mainly during the northern sample observations) prevented us from reaching this goal. For this reason the  final WINGS-SPE sample comprises $48$ (of the 77) clusters, $22$ of which are in the southern sky and $26$ are in the northern sky. This sample was extracted from the main sample giving the highest observing priority to clusters with few (less than 20) or no redshifts available from the literature. For example, clusters A0085, A0548b, and Abell 3558, which have large databases of available redshifts, had a lower priority. 
 
The target selection was based on the available WINGS optical B, V photometry (\cite{varela08}) and the aim of the target selection strategy was to maximize the chances of observing galaxies at the cluster redshift without biasing the cluster sample. The main criteria for selecting spectroscopic targets were similar for the northern and southern samples, the only difference being the fibre size. In both cases, we selected galaxies with a V magnitude within the fiber aperture of $V < 21.5$ (although in a very few cases galaxies at fainter magnitudes have been observed) and with a color within a 5kpc aperture of $(B-V)_{5kpc}\lesssim 1.4$. These loose selection limits were applied so as to avoid any bias in the observed morphological type, as is the case of a selection based on the CM relation only (which selects just red, early type galaxies). The exact cut in color was varied slightly from cluster to cluster in order to account for the redshift variation and to optimize the observational setup. Our total apparent magnitude limit ($V \sim 20$) is 1.5 to 2.0 mag deeper than the 2dFRS and Sloan surveys, respectively, and this is, in general, reflected in a higher mean number of member galaxies detected per cluster.

\addtocounter{table}{1}
In Table \ref{table:prop}, we list the main properties of the clusters in the WINGS-SPE sample. The different columns indicate: (1) cluster name, (2-3) coordinates of the image field center at epoch $J2000$ [right ascension (2) in hours, minutes and seconds and declination (3) in degrees, arcminutes, arcseconds], (4) number of redshifts, which is equal to the
number of entries in the spectroscopic catalog for that cluster, (5) mean cluster redshift, (6) number of member galaxies (used to compute mean redshift and velocity dispersion as explained in \S 4), (7) cluster velocity dispersion computed from the WINGS-SPE data, (8) cluster velocity dispersion available from the literature, (9) reference for the literature velocity dispersion, (10) logarithm of the X-ray luminosity in the $0.1-2.4$\,keV ROSAT RASS bandpass (from \cite{ebeling96}), (11) virial radius in $Mpc$, (12) aperture in units of $R_{200}$. For completeness, we also list in Table \addtocounter{table}{1}\ref{table:prop2} the main properties of the 30 clusters not observed but for which literature data exist.  The different columns here indicate: (1) cluster name, (2-3) coordinates of the image field center at epoch $J2000$ [right ascension (2) in hours, minutes and seconds and declination (3) in degrees, arcminutes, arcseconds], (4) cluster mean redshift, (5) number of member galaxies, (6) cluster velocity dispersion, (7) reference for the data reported in columns (4-6), (8) logarithm of the X-ray luminosity in the $0.1-2.4$\,keV ROSAT RASS bandpass (from \cite{ebeling96}).

\section{Spectroscopic data}

\subsection{Observations and Data Reduction}
The spectroscopic observations were obtained over the course of 6 observing runs (22 nights) at the 4.2\,m William Herschel Telescope (WHT) using the AF2/WYFFOS multifiber spectrograph and 3 observing runs (11 nights) at the 3.9\,m Anglo Australian Telescope (AAT) using the 2dF multifiber spectrograph. The spectroscopic runs are summarized in Table \ref{table:runs}. 

AF2/WYFFOS is the multi-object, wide-field, fiber spectrograph working at the 4.2m
William Herschel Telescope (WHT) on La Palma.  AF2/WYFFOS can allocate up to 150 fibres, each of 1.6\,arcsec diameter.  We allocated typically 60-70 fibers to objects in a given configuration, and on average, 15-20 sky fibers. For runs $1-5$, we used the AF2/WYFFOS Long Camera with the TEK6 1024$\times$1024 CCD with pixel size of 24 $\mu$. In combination with the 600B grism that was used, this yielded a spectral resolution of $\sim$ 6 \AA~ FWHM, depending on the location on the CCD.  For run $6$ we again used the 600B grism, but with the  2-chip EEV $4300\times4300$ mosaic with 13.5 $\mu$ pixels. With a 2x2 binning of the CCD pixels, this gave a spectral resolution of $\sim$ 3 \AA~ FWHM. The spectra were centred at a wavelength of 5100\AA\ and covered the range $\sim$3800--7000\AA. Hence they covered many interesting spectral features ranging from Ca H\&K in the blue to NaD in the red. The galaxies were divided into different configurations, depending on their luminosities. Two configurations were executed for each cluster: one bright ($V<20.5$ inside the 1.6$"$ aperture) and one faint ($20.5<V<21.5$ inside the 1.6$"$ aperture), with total exposure times of 1\,hour for the bright sample and 2\,hours for the faint one. He/Argon lamp exposures  for wavelength calibration and offset sky exposures for fibre throughput calibration were also obtained.  We also observed spectroscopic flux standards, Lick standards, and radial velocity standards during twilight.  The reduction of these multi-fiber spectra was performed using the {\it dofiber} IRAF package.  
\begin{table}[h!]
\centering
\caption{The WINGS-SPE observing runs}
\begin{tabular}{clrc}
Run & Telescope& Date~~~~ &  $\lambda$ FWHM ($\AA$)\\ 
\hline \\
1 & WYFFOS@WHT & Aug/Sept 2002 &  6 \\
2 & WYFFOS@WHT & April 2003 &  6 \\
3 &WYFFOS@WHT & June 2003 &   6 \\
4 & WYFFOS@WHT & March 2004 &   6 \\
5 & WYFFOS@WHT & June 2004 &   6 \\
6 & WYFFOS@WHT & Oct. 2004 &  3.2 \\
7 & 2dF@AAT & Jan. 2003 &  9 \\
8 & 2dF@AAT & Sept. 2003 &   9 \\
9& 2dF@AAT & March 2004 &   9 \\
\hline
\label{table:runs}
\end{tabular}
\end{table}

The multiple exposures in each pointing were combined, with cosmic rays being rejected in the process.
Twilight sky flats or combined object frames were used to define the apertures and trace the spectra
on the CCD.  The median offset sky exposures were then used to calculate
the throughput for each fiber, and to normalize all of the sky fibres. The arc spectra were extracted and matched with standard arc lines to determine the dispersion solution using a polynomial fit. 
These fits yielded a typical rms scatter of 0.05 \AA. Finally, the object spectra were extracted, normalized, and wavelength calibrated.  A ÔÔmaster ÕÕ sky spectrum was derived for each exposure by 
combining the spectra from $10-30$ individual sky fibers. The fibers do not all have identical throughput, and in some runs the differences could not be adequately determined from the calibration ÔÔflat-fieldÕÕ spectra obtained. In order to perform accurate sky subtraction, we scaled the master sky spectrum based on the flux in the bright 5577 \AA~ line to minimize residuals in each galaxy spectrum.  At the end of this procedure the sky subtraction accuracy was quite good, ranging between 1-3\% (defined as the rms of the normalized sky fibres about the master sky spectrum). 
\begin{figure*}
\centering\includegraphics[scale=0.8]{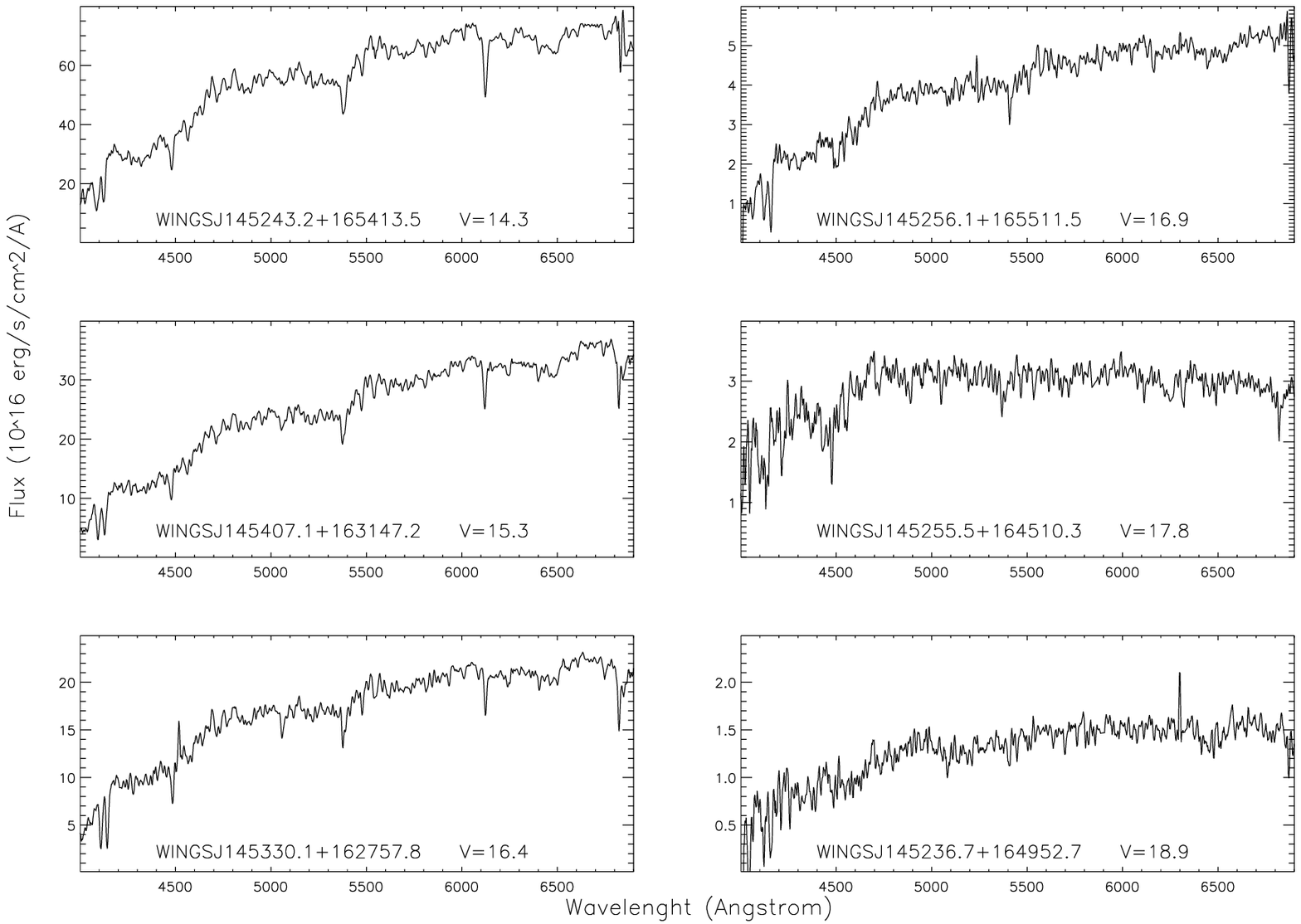}
\caption{Sample spectra for the northern sample observed with
WYFFOS spectrograph at WHT. Spectra of bright (left) and faint (right) 
galaxies are
shown for the cluster A1983. The object IDs and V-band magnitudes are given in each panel.}
\label{fig:A1983_spec}
\end{figure*}

\begin{figure*}
\centering\includegraphics[scale=0.8]{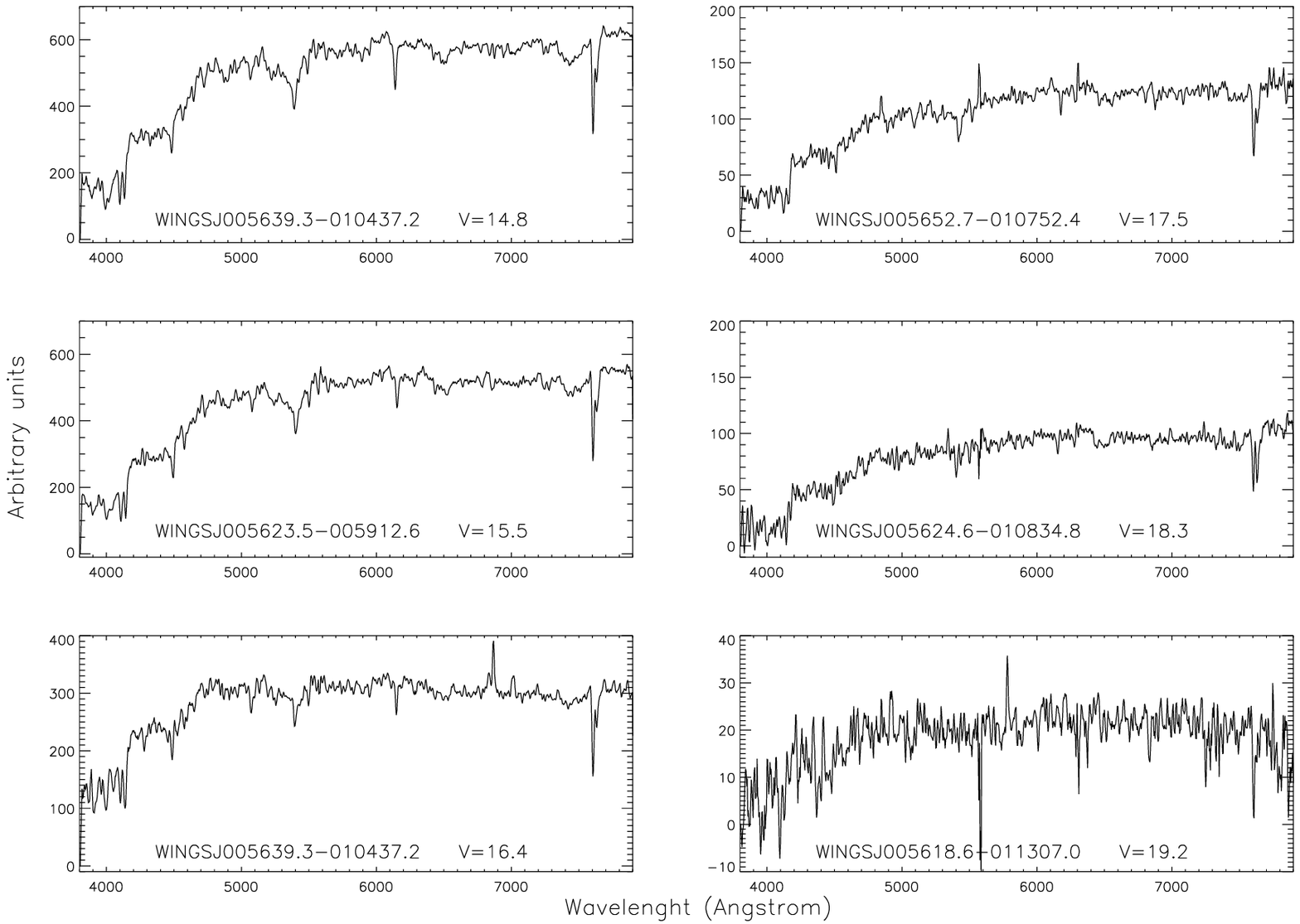}
\caption{Example of spectra for the southern sample observed with 2dF
spectrograph at AAT. Spectra of bright (left) and faint (right) galaxies 
are shown for
the cluster A0119. The object IDs and V-band magnitudes are given in each panel.}
\label{fig:A0119_spec}
\end{figure*}

For the southern sample, we used the Two-degree Field (2dF) multifiber spectrograph on the AAT (runs 7-9). This instrument can observe up to 400 objects simultaneously over a two degree field of view. The detectors were 1024$\times$1024 24$\mu$m pixel Tektronix CCDs, which in combination with the 300B grating yielded a resolution of $9\,\AA$ and a wavelength range of  $3600-8000\,\AA$. The fiber diameter is $2\arcsec$. The galaxies were again divided into two different configurations in order to observe multiple sets galaxies brighter and fainter than
$V=19.5$ in the fibre. For each cluster we were able in this way to observe $\sim 150-200$ target galaxies. The integration times were generally $1$ and $2$ hours for the bright and faint configurations, respectively. For each field we first took a multi-fiber flat-field exposure using the quartz lamp in the calibration unit. This flat-field is used to trace the positions of the fibres on the CCD image, to fit the spatial profile of each fiber as a function of wavelength, and to apply a 1-dimensional pixel-to-pixel flat-field correction to the extracted spectra. Exposures of helium and copper-argon arc lamps were taken for wavelength calibration.  The data were reduced at the telescope using the 2dF data reduction ({\it 2dfdr}) pipeline software, a full description of which is given by \cite{Taylor96} (see also http://www.aao.gov.au/2df). The main steps in the process are as follows: bias subtraction and flat fielding, mapping of the spectra with background subtraction and finally wavelength calibration. However the standard  sky subtraction does not work well for all the spectra in general systematic residuals are evident where skylines have been subtracted, in particular for the faint ones. For this reason we have re-extracted from each configuration the spectra from individual sky fibers, derived a master sky spectra and subtracted this spectrum from the original wavelength calibrated spectra. At the end of this procedure the sky subtraction accuracy was quite good, ranging between 1-3\% (as in the case of the northern sample). Sample spectra of galaxies with different luminosities in the bright and faint samples are shown in Figure \ref{fig:A1983_spec} for the northern sample and in Figure \ref{fig:A0119_spec} for the southern one.

\subsection{Flux calibration}
Absolute flux calibration can never be done with fibers, since we are limited by the
fixed fiber diameter. Following the recipes of the 2dFGRS survey  (Lewis et al. \cite{lewis}), the following procedure was adopted to perform a relative flux calibration. First we applied the response function available from 2dF web site (Lewis et al. \cite{lewis}). This response curve can be applied, on average, to give an approximate relative flux calibration for the 2dF spectra. However, the results for individual spectra will vary considerably due to sky subtraction and efficiency variations over each plate. In order to obtain an optimal flux calibration correction we decided to perform a comparison with SDSS spectra for a set of galaxies in common to the two samples. From this comparison we derived a mean correction curve that was applied to all the spectra of the southern sample. 

For the northern sample, spectrophotometric standard stars were
observed in order to be able to flux calibrate the data. These star
spectra were reduced and wavelength calibrated with the same
\verb"dofiber" package. Typically, one star was observed at the start
of the night and another one at the end of the night and at least in
two different fibers. For the nights for which a
full sample of standard stars in different fibers was available, we used
the average sensitivity function derived using all stars observed that
night. Unfortunately, due to bad weather conditions
there were nights when each flux standard was observed through one
fiber only. Having verified that the curve from fiber to fiber for the 
different standard stars all agreed well (less than 10\% difference),
the data taken in those nights lacking a full sample of standard stars
were calibrated using the average sensitivity function of all the
other nights.
As a final step, all the spectra were corrected for atmospheric extinction using the extinction curves published for Siding Spring and La Palma Observatories, respectively.   

\section{Redshifts Measurements}
In this work, radial velocities were measured using procedures based on the Fourier cross-correlation method (Tonry \& Davis \cite{tonry}), as implemented in the {\it xcsao} task in the RVSAO package. RVSAO (\cite{KM}) is an IRAF add-on package developed at the Smithsonian Astrophysical Observatory Telescope Data Center to obtain radial velocities from spectra using cross-correlation and emission line fitting techniques.

We note that the redshifts obtained via {\it xcsao} are unreliable in cases where the observed spectrum departs markedly from that of the template star (taken from \cite{J84}), or when the galaxy spectrum contains strong emission lines. To determine redshifts with greater robustness, a set of template galaxy spectra was also used. As template galaxy spectra we used a sample of spectra with high S/N extracted from our data as well as synthetic emission line spectra 
generated using the task {\verb"RVSAO/linespec"}. This task reads a list of positions of emission lines and creates a spectrum with Gaussian lines of the indicated half-widths at the indicated positions, writing a one-dimensional IRAF file. The galaxy templates include both pure absorption and emission-dominated spectra. All the spectra were inspected by eye to verify the redshift measurements and to ensure that the best matching template was chosen. Finally we checked the spectra for possible residual shifts due to systematic errors in the wavelength calibration. In order to correct for residual shifts we cross-correlated each spectrum (before sky subtraction)  with a template sky spectrum. The final redshift has been corrected for the measured displacements of the skylines  with respect to the zero-point. After this correction, each radial velocity measurement was corrected to the heliocentric velocity. We chose to limit the spectral range used in the cross-correlation to the interval $3800\,\AA\leq\lambda\leq 6800\,\AA$, because in the blue region the spectra are very noisy outside of this range, and in the red region the spectra are dominated by the strong telluric bands. 
\begin{figure*}
  \includegraphics[scale=.345]{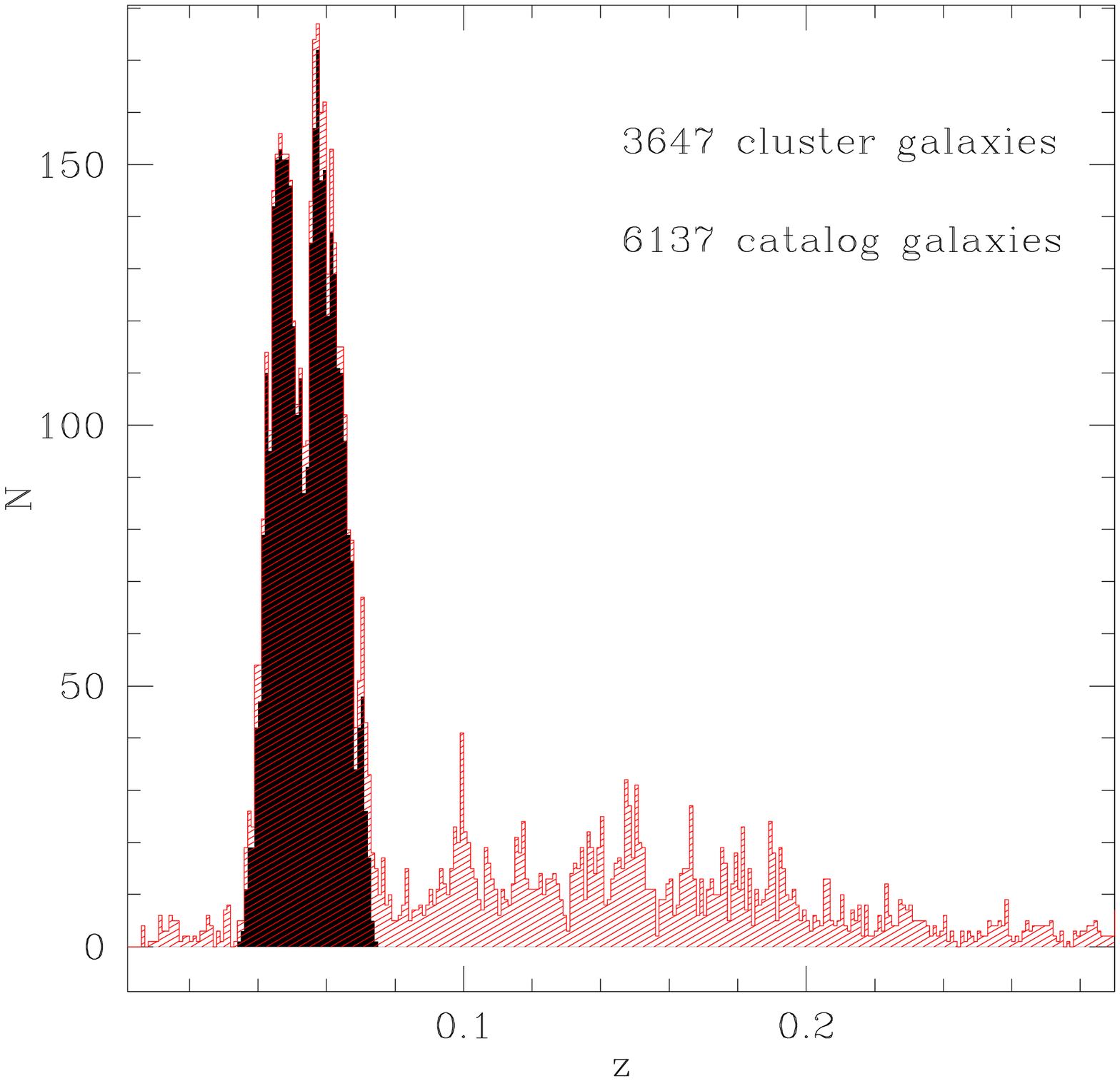}  \hspace{2truecm}
  \includegraphics[scale=.345]{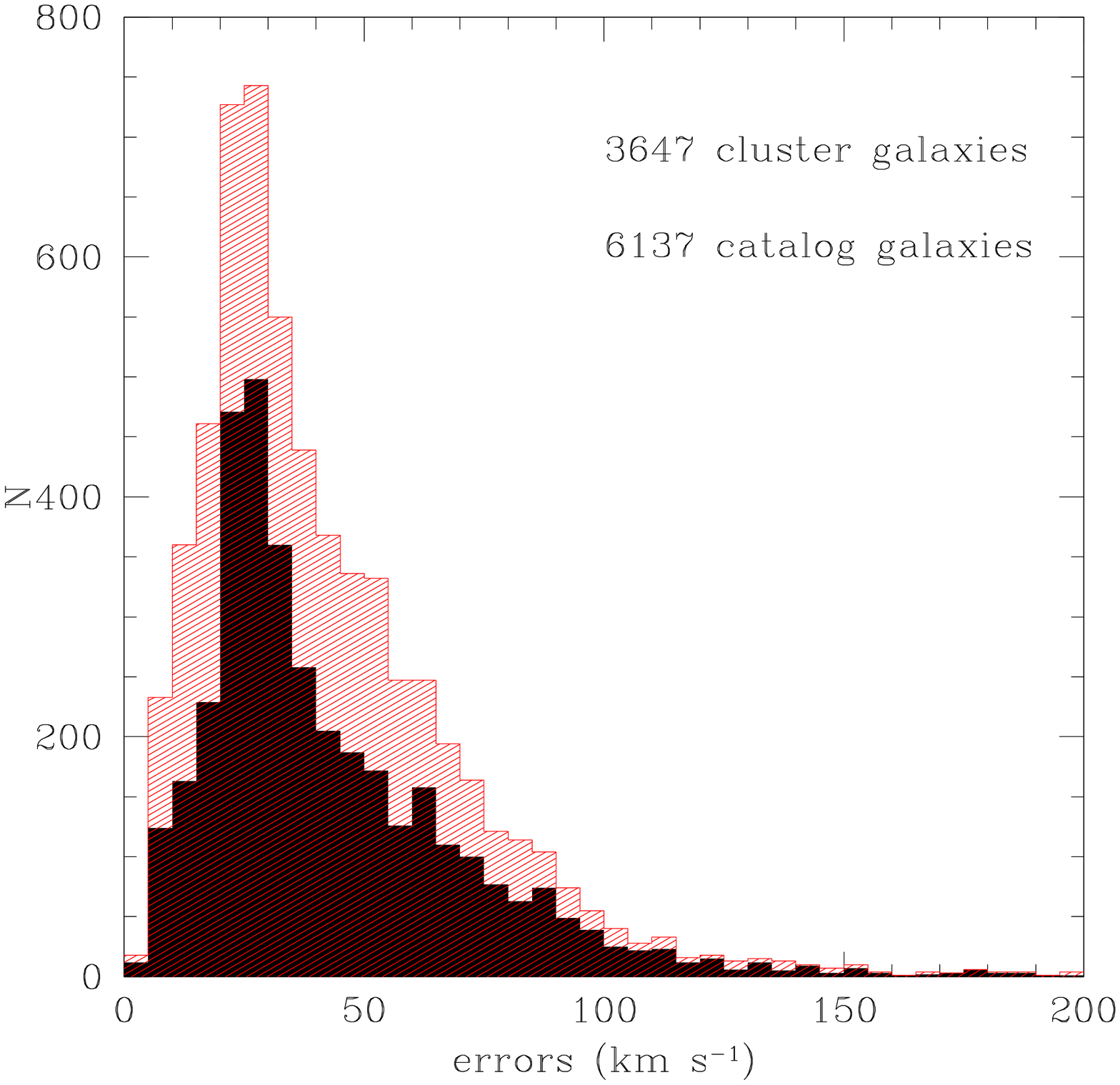}  
  \caption{Distribution of redshifts (left panel) and errors (right panel) over the complete WINGS sample. Red-shaded histograms show the overall WINGS redshift distribution, the black histograms show only the
distribution of galaxy members.}\label{fig:tot_distr}
\end{figure*}

\begin{table*}[h]
\begin{center}
\caption{Example of data table: see online material for the complete version}
\begin{tabular}{lccccccc}
Name & $R.A.$ & DEC & cz & err & $r$ & Memb. &$Field$ \\ 
 & ($J2000$) & ($J2000$) & ($kms^{-1}$) &  ($kms^{-1}$) & & flag &\\
\hline
WINGSJ005623.55-005912.6 & 00:56:23.558 & -00:59:12.665 & 13230 & 28 & 11.0 & 1&A0119  \\
WINGSJ005544.38-005926.6 & 00:55:44.385 & -00:59:26.631 & 29765 & 48 & 4.1 & 0 & A0119 \\
WINGSJ005552.84-005935.7 & 00:55:52.844 & -00:59:35.776 & 51996 & 22 & 6.6 & 0 &A0119  \\
WINGSJ005655.95-005948.2 & 00:56:55.951 & -00:59:48.256 & 13893 & 45 & 5.5 & 1 & A0119  \\
WINGSJ005540.86-005949.5 & 00:55:40.862 & -00:59:49.588 & 13419 & 40 & 5.7 & 1 & A0119  \\
WINGSJ005612.25-010005.2 & 00:56:12.251 & -01:00:05.207 & 52543 & 17 & 9.0 & 0 & A0119 \\
WINGSJ005651.59-010025.7 & 00:56:51.599 & -01:00:25.791 & 14347 & 37 & 6.5 & 1 & A0119  \\
WINGSJ005620.31-010022.6 & 00:56:20.310 & -01:00:22.628 & 48308 & 58 & 2.5 & 0 & A0119  \\
WINGSJ005631.67-010044.9 & 00:56:31.675 & -01:00:44.997 & 42181 & 92 & 4.0 & 0 & A0119 \\
WINGSJ005548.42-010043.1 & 00:55:48.424 & -01:00:43.168 & 44006 & 25 & 5.2 & 0 & A0119  \\
........ & ........ & ......... & ........ & .... & .... & .... &.....\\
\hline
\end{tabular}
\end{center}
NOTES: the columns indicate $(1)$ galaxy names, fiber $J2000$ positions, $(2)$ right ascension (in hours,minutes and seconds), $(3)$ declination (in degrees, arcminutes, arcseconds), (4) heliocentric redshift (in $kms^{-1}$), (5) error on the redshift measurement (in $kms^{-1}$), (6) cross-correlation factor, (7) membership flag and (8) cluster field.
\label{table:catalogs}
\end{table*} 
Summarizing, the procedure followed for measuring the redshifts was as follows:
\begin{itemize}
    \item   determination of the redshift of the galaxies using a semi-automated method;
    \item   determination of the displacement of the skylines with respect to the zero-point and correction for this;
    \item   spectra with uncertain redshifts flagged and checked again manually to ensure the best reliability.
\end{itemize} 

\noindent
The redshift distribution of the whole spectroscopic sample is presented in Fig \ref{fig:tot_distr}. An example catalogue can be seen in Table \ref{table:catalogs}. 
The different columns refer to:\\
column (1) gives the object name,  \\
column (2) gives the right ascension (J2000), and column (3) gives the declination (J2000), \\
column (4) gives the redshift, in km\,sec$^{-1}$, and column (5) the redshift error in km\,sec$^{-1}$, \\
column (6) gives the correlation factor $r$ defined as:
\begin{equation}
r=\frac{h}{\sqrt{2}\sigma_a}
\end{equation}
which is the ratio between the height, $h$, of the main peak to the mean height, $\sigma_a$, of the secondary peaks in the Fourier cross-correlation function (Tonry \& Davies \cite{tonry}). Assuming sinusoidal noise, with the half-width of the sinusoid equal to the half-width of the correlation peak this factor can be related to the measurement error (\cite{KM}) as:
\begin{equation}
\delta=\frac{3}{8}\frac{w}{1+r}
\end{equation}
where $w$ is the FWHM of the correlation peak. Only spectra achieving a correlation factor higher than $2$ have been considered reliable after the visual check, the mean $r$ is equal to $8$ and $83\%$ of the redshift determinations have a correlation factor greater than $4$ ensuring the measurements are highly reliable. \\
Column (7) gives the membership flag; a value of 1 indicates the galaxy is a cluster member, and a value of 0 indicates it is either a background or foreground galaxy. Cluster membership is defined on the basis of the galaxy redshift being within $\pm3\sigma$ from $z_{cl}$ (see \S 6).\\
Column (8) indicates the cluster field to which the galaxy belongs.
The complete version of Table~\ref{table:catalogs} is available in electronic format only. 

\section{Data quality}

\subsection{External comparisons}

Previous spectroscopic surveys have yielded redshift  measurements for a substantial number of galaxies in common with the WINGS-SPE sample. Here we employ these data to test the reliability of our data. We use data from the NASA/IPAC Extragalactic Database (NED) to perform our comparison with the aim of extending our catalogs and exploiting the entire data-set (WINGS+literature) in the kinematical and dynamical analysis. In particular, we intend to use these data to study in detail the properties of clusters and substructures in the WINGS sample (see \cite{ramella07} for the study of substructures in 2D). We also compare separately our data with the NOAO Fundamental Plane Survey (NOAO-FPS, \cite{noaoI}), which has a good overlap with our sample, and with the SDSS with which we only have 12 clusters in common.
These three comparison samples allow us to compare the data with a uniform (NOAO-FPS and SDSS) and a non-uniform but more extended catalog (NED) in order to check the data quality and also to complete and extend our catalogs with external data.

\begin{figure}
    \includegraphics[scale=.4]{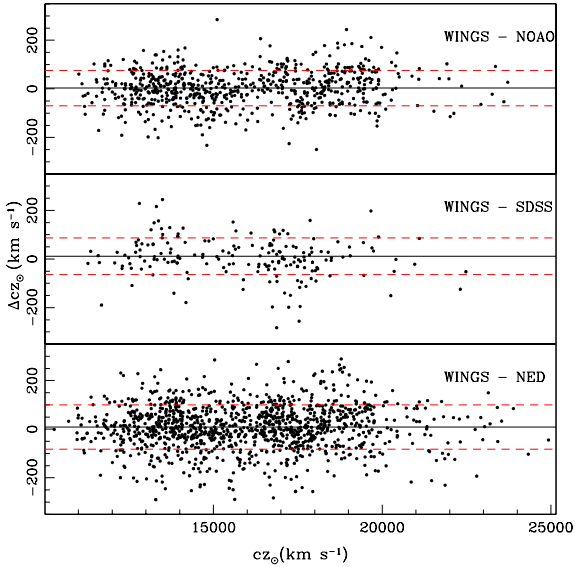}
  \caption{Plot of the residuals for redshift measurements obtained for the $3$ comparison samples used to perform the quality check: NOAO-FPS, SDSS, NED (from top to bottom). The black line indicates the offset while the red dashed lines indicate the scatter.}\label{fig:comp2}
\end{figure}
 
To make the comparison we selected data in common clusters, cross-correlating the catalogs and taking as common the galaxies whose coordinates differ by less than $6$ arcsec. Few galaxies had redshifts that differed by more than $300$\,km\,s$^{-1}$. We checked these galaxies one by one, finding that in all cases the difference arose from a mismatching in the catalogs, so we discarded these objects from the comparison sample. The application of the selection criteria leave us with a final comparison sample of $2,218$ galaxies.  For the 31 clusters in common with the NOAO-FPS, we have a ratio $R=N_{wings}/N_{noao-fps}=1.73$ of measured redshifts, were $N_{wings}$ is the number of galaxies with redshifts in the WINGS sample and $N_{noao-fps}$ is the number for the NOAO-FPS survey. In particular, this ratio is $R=N_{wings}/N_{noao-fps}=0.88$ for the northern sample (14 clusters) and $R=N_{wings}/N_{noao-fps}=2.34$ for the southern sample (16 clusters), emphasizing the different completeness levels in our two subsamples (north and south). In Figure \ref{fig:comp2} we show the global comparison for the $1,325$ galaxies in common with NED, the 217 in common with SDSS, and $676$ in common with NOAO-FPS.

As summarized in Table \ref{table:comp}, the mean differences between WINGS and the data in the literature are very low and much lower than the dispersion assuring the absence of systematic offsets. The different columns refer to: (1) comparison sample name, (2) mean difference (in km\,s$^{-1}$), (3) rms scatter in the differences, and (4) number of galaxies used in the comparison.

\begin{table}[h!]
\caption{External comparison of redshift}
\begin{tabular}{lccc}
\hline
Sample & Offset & Scatter & N \\
& $kms^{-1}$ & $kms^{-1}$ & \\
\hline
WINGS--NOAO-FPS & $3\pm3$ & $70$ & $676$\\
WINGS--SDSS & $11\pm5$ & $75$ & $217$\\
WINGS--NED & $8\pm3$ & $90$ & $1108$ \\
\hline\hline
\label{table:comp}
\end{tabular}
\end{table}

Moreover, the dispersion in the measurements is low enough to not greatly influence the measurement of the internal velocity dispersion of galaxy clusters, even in the cases where this quantity is considered low (300-400\,km\,s$^{-1}$) as, for example, for substructures and groups. This fact is fundamental in view of the subsequent dynamical analyses. Particularly discrepant cases can be checked and corrected cluster by cluster.

In addition to the three large comparison samples presented above, we have also considered the smaller sample of galaxy clusters presented in the previous work by \cite{bett06}. There are 3 clusters with a total of 23 galaxies in common with this data sample. The mean difference ($\Delta$cz) is $\sim -9 \pm 24$\,km\,s$^{-1}$, while the rms scatter in this case is $\sim 118$\,km\,s$^{-1}$. The larger scatter is due mainly to the small number statistics. 

\subsection{Completeness and success rate}

It is very important to know the completeness level of the spectroscopic
observations as this is a factor that must be accounted for in the derivation of
luminosity functions, M/L ratios, as well as anytime we want to use the spectroscopic
sample to study magnitude-dependent properties (e.g. the different galaxy
population fractions inside clusters, see Poggianti et al. \cite{poggianti06}). Following the selection criteria described in Section \ref{sec2}, the exact cut in the color-magnitude diagram varied slightly from cluster to cluster (ranging in the interval $1.2\lesssim(B-V)_{5kpc}<1.4$) due to the small differences in cluster redshift and to minimize the level of contamination from the background. In a few cases, the cut has purposely included  a secondary red sequence, such as for Abell 151, to be able to study also background clusters. The completeness as a function of magnitude is defined here as:
\begin{equation}\label{eq:compl}
C(m)=\frac{N_{z}}{N_{ph}}(m)
\end{equation}
where $N_z$ is the number of galaxies with measured redshifts and $N_{ph}$
is the number of galaxies in the parent photometric catalog, taking into
account the cuts in color and magnitude, for each
given magnitude bin $m$. Completeness is usually a decreasing function of the magnitude because in observations priority is given to brighter objects. The success rate, that is the fraction of galaxies with redshift determination with respect to the total number of observed galaxies, is similarly defined as:
\begin{equation}
SR(m)=\frac{N_{z}}{N_{tg}}(m)
\end{equation}
where $N_z$ is defined as in eq. \ref{eq:compl} and $N_{tg}$ is the number
of target galaxies we actually observed.

\begin{figure}
  \includegraphics[scale=.37]{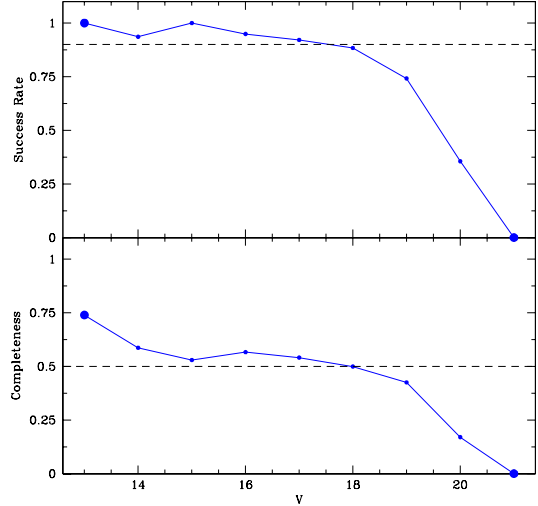}
  \caption{Success rate (upper panel) and global completeness (lower panel)
for the whole WINGS-SPE sample. Big dots represents bins with low statistic
(less than $20$ points).}\label{fig:compl}
\end{figure}

\begin{figure*}
\includegraphics[scale=.37]{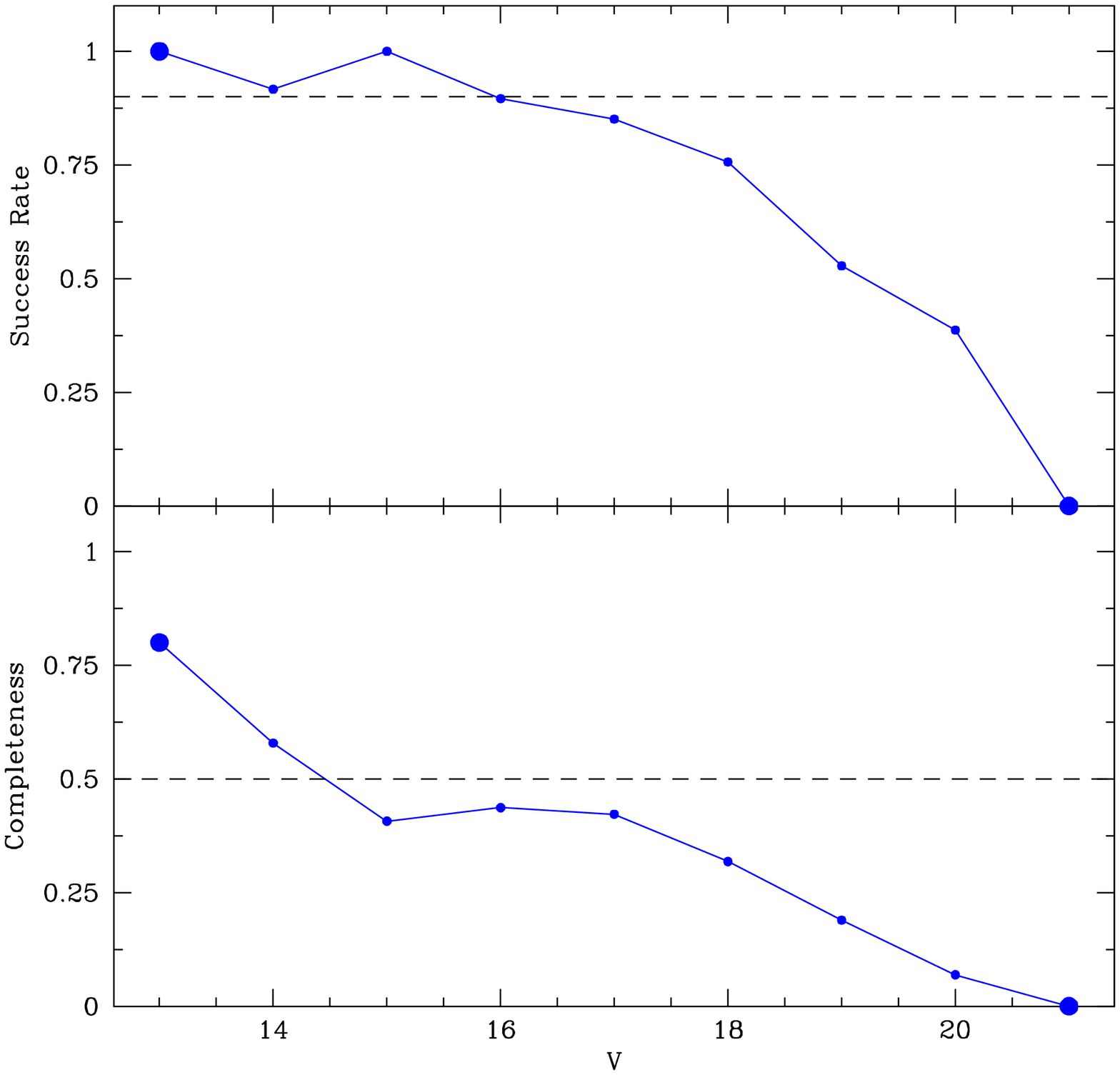} \hspace{2truecm}
  \includegraphics[scale=.37]{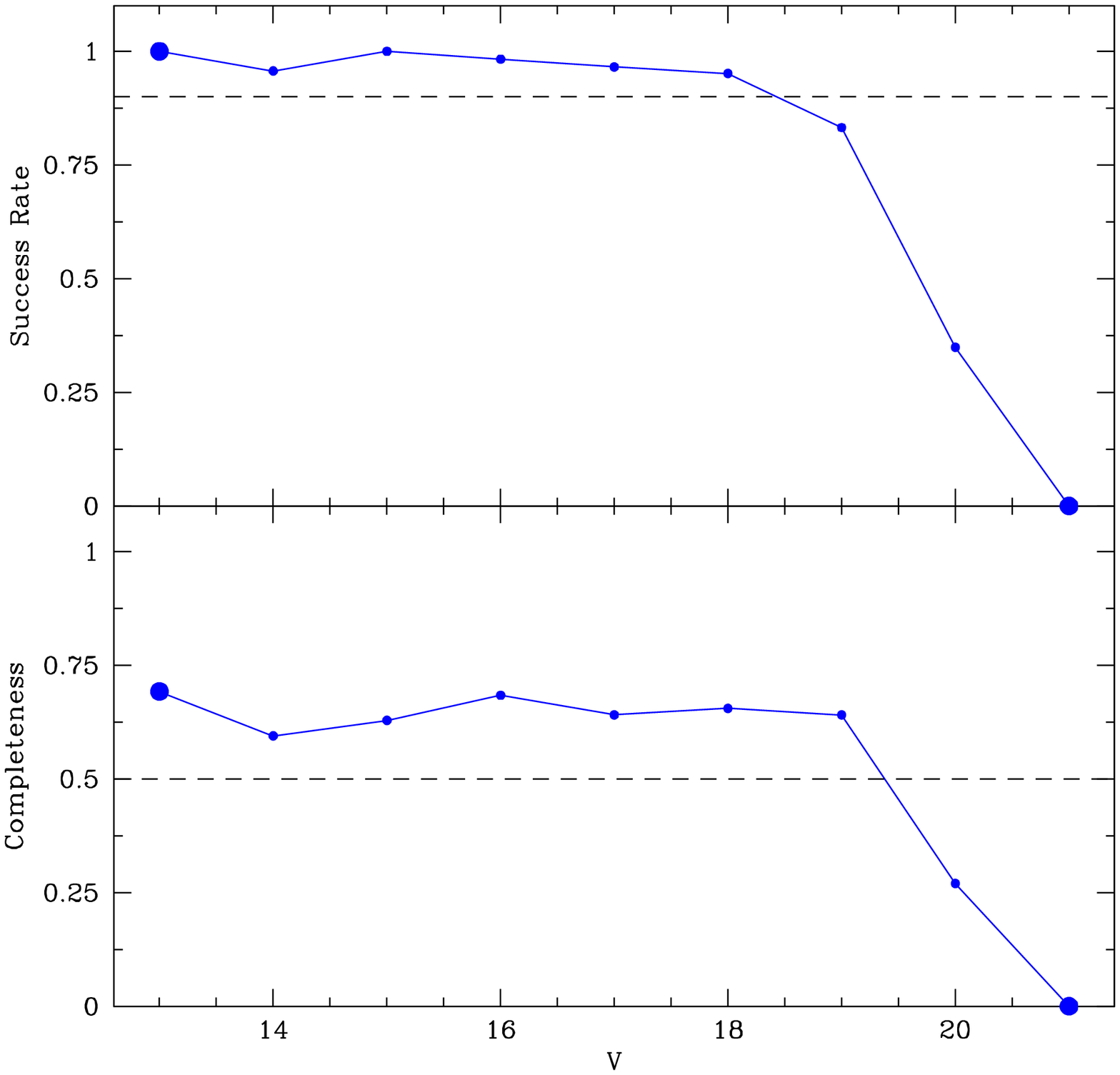}
  \caption{Success rate and global completeness for WYFFOS (left plot) and
2dF observations (right plot). Big dots represents bins with low statistic
(less than $10$ points).}\label{fig:compl2}
\end{figure*}

The success rate and completeness as a function of $V$ magnitude are shown for the 
WINGS-SPE sample in Fig.\ref{fig:compl}. We also computed these
two functions separately for the two  subsamples observed with the 2dF and
WYFFOS spectrographs; they are shown in Fig. \ref{fig:compl2}. We want to emphasise that the quite large difference in the completeness and success rates for the two sub-samples is mainly due to two reasons: first of all, during the observations of the northern sample we lost $\sim$30\% of the observing time due to bad weather conditions. Secondly, the large difference in the number of fibers available (150 with WYFFOS and 400 with AF2) and hence the multiplex power of the two instruments. The effect of the bad weather on the northern sample is particularly evident in the upper panels of Fig. \ref{fig:compl2}. Many galaxies at fainter apparent magnitudes ($V\gtrsim 18$) are lost because of the fact that some faint configurations have spectra with very low S/N and were completely unusable. At variance with this situation, the southern observations show a very flat behavior up to the magnitude limit of the observations (V$\sim 19.5$). Respectively, looking at the completeness (lower panels in the figure), the worsening effect due to the bad weather is strengthened by the difference in the number of available fibers that could be placed using the two spectrographs. The almost constant behavior of the completeness for the 2dF observations has to be compared with the resulting monotonically decreasing completeness of the WYFFOS data.


\section{Results}
\subsection{Cluster Assignment and redshift histograms}
The modelling of galaxy kinematics in clusters remains one of the major tools in determining
cluster properties, in particular their mass distribution and dark matter content.
Due to projection effects, any cluster kinematic data sample inevitably contains
galaxies that are not bound to the cluster and therefore are not good tracers of its gravitational
potential. These galaxies are called interlopers. An essential step in dynamical
modelling of clusters by any method is therefore to remove such interlopers from the
samples or take their presence into account statistically.\\
The peculiar velocity of a galaxy with redshift z in the rest-frame of a cluster with redshift $z_{cl}$ is given by
\begin{equation}\label{zeq}
v = c\frac{z - z_{cl}}{1 + z_{cl}}
\end{equation}
valid to first-order for $v\ll c$ (e.g. \cite{H74}, \cite{carlberg96}). The dispersion of the $v$ values for the cluster members define the cluster rest-frame velocity dispersion
$\sigma_{cl}$ that is related to the observed velocity dispersion:
\begin{equation}
\sigma_{cl}=\frac{\sigma_{obs}}{1+z_{cl}}.
\end{equation}
It is very important to determine as accurately as possible which galaxies belong to the clusters and which ones have to be considered as interlopers, and thus removed. In fact, it is known that the presence of interlopers can increase the value of the observed velocity dispersion and since the estimated virial mass is proportional to the third power of this value, a small error in the velocity dispersion can highly influence the mass estimate. We employ an iterative $\pm3\sigma$ clipping scheme to determine which galaxies are cluster members (Yahil and Vidal, \cite{YV}). This works as follows. 

A first estimate of $z_{cl}$ is obtained from a visual inspection of the redshift histogram, usually corresponding to the statistical mode for the given distribution. Galaxies with redshifts outside the region $z_{cl}\pm0.015$ (corresponding to $\sim 4000$\,km\,s$^{-1}$) are removed and not used in any further analysis. The following two steps are then iterated until convergence on $z_{cl}$ and $\sigma_{cl}$ is reached: (1) calculate $v$ for all the galaxies using Eq.\ref{zeq}; (2) for galaxies with $v$ in the interval [$-3\sigma_{cl}$,$+3\sigma_{cl}$], a new estimate of $z_{cl}$ and $\sigma_{cl}$ is calculated using the robust biweight location and scale estimators (Beers et al. 1990).

This approach is still widely used today (e.g., see \cite{MJ08}) even if different methods of interloper removal based on dynamical or statistical restrictions imposed on ranges of positions and velocities available to cluster members have been developed. We have decided to use this method because, despite of its intrinsic simplicity, it has been demonstrated to be the most effective in determine cluster membership in many cases (e.g., \cite{woj07}). Since in this paper we are only interested in the global value of the cluster velocity dispersions, we do not  need here to investigate different methods of interloper removals, as the values of the cluster velocity dispersions would only marginally be modified. We will perform a more detailed analysis of the dynamics of the WINGS galaxy clusters in a following paper (Cava et al., in prep.), where more sophisticated methods of interloper removal will also be exploited. 

 Here we want to remark upon the fact that our velocity dispersions are not all evaluated at the cluster virial radius, since our observations do not always go that far. Cluster velocity dispersions are known to depend on radius, hence on the aperture used to measure them (\cite{Fadda}; \cite{Rines}; \cite{Muriel}; \cite{GM01}; \cite{aguerri07}). However, the dependence is very mild for galaxy clusters in the range of apertures considered here, i.e. $(\sim\,0.8\,\pm\,0.2)\times R_{200}$, and probably smaller than 10\%, as demonstrated for example by the analysis of \cite{Lokas} (see their Fig.7) or as shown recently by \cite{aguerri07} for a large sample of nearby clusters (they find a variation lower than $3\%$ recomputing the velocity dispersions inside apertures of $0.4\times\,R_{200}$ and $0.6\times\,R_{200}$). In columns $11$ and $12$ of Table \ref{table:prop} we report  the virial radius in  $Mpc$ and the maximum observed aperture radius in unit of $R_{200}$ for each cluster, that is the maximum dynamical radius for which we have spectroscopic data. From Table 1 can also be seen that $\sim\,40\%$ of the observed clusters achieve an aperture radius $\gtrsim 0.9\times R_{200}$.

Velocity dispersions estimates, $\sigma_W$ (in km\,s$^{-1}$), for all the WINGS-SPE sample are listed in column $7$ of Table \ref{table:prop}. The errors quoted here were obtained using the classical jackknife technique (\cite{E82}). For comparison, values of $\sigma$ found in the literature ($\sigma_L$) are given in column $8$.

\begin{figure*}
  \includegraphics[scale=.8]{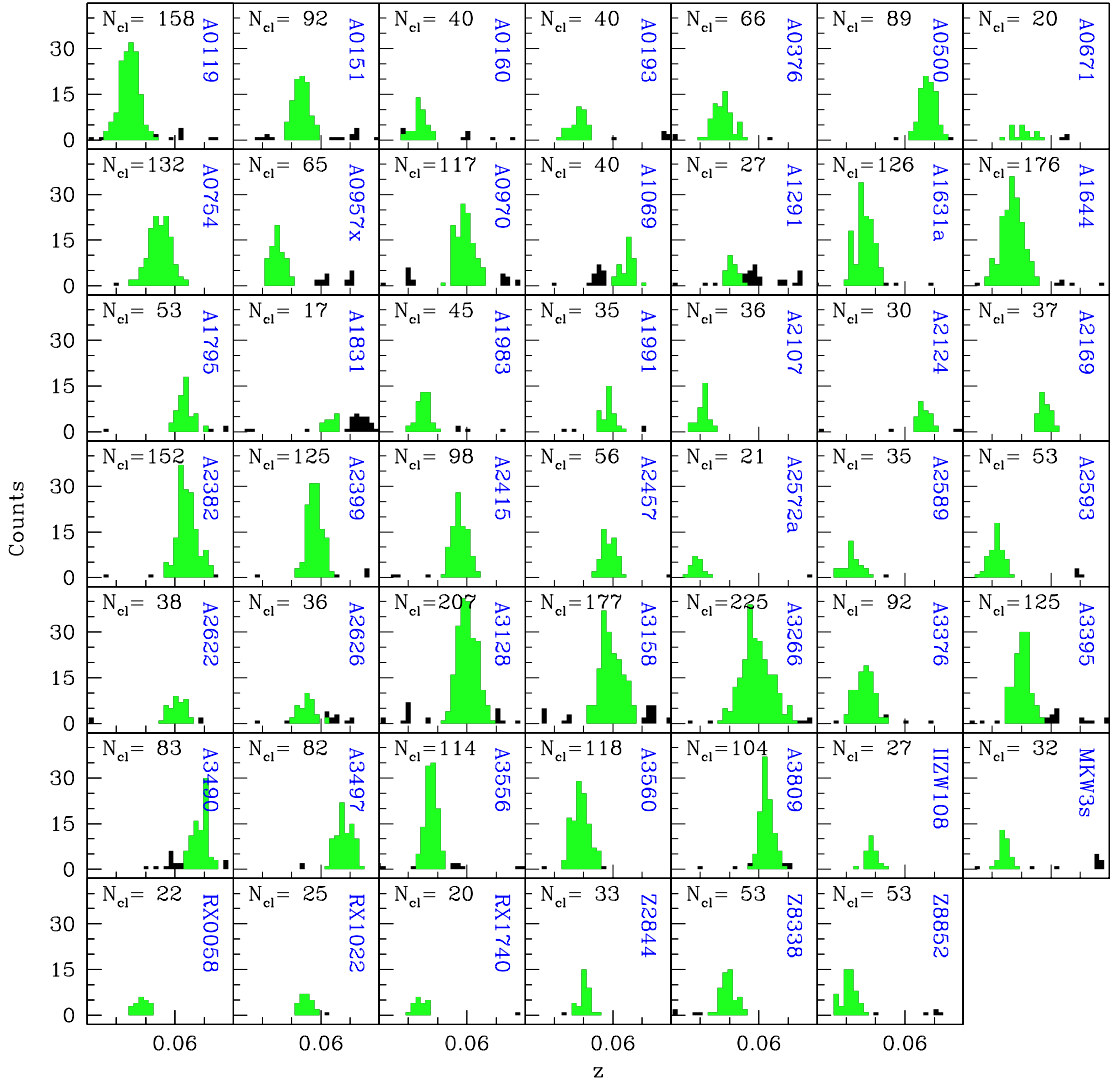}  
  \caption{Histograms of the WINGS-SPE data sample in the range $0.03\le z \le 0.08$. In color are histograms for cluster members defined using a $3\sigma$ cut. $N_{cl}$ indicates the number of cluster galaxies inside $3\sigma$ from the mean cluster redshift.}\label{fig:hist1}
\end{figure*}

\begin{figure*}
  \includegraphics[scale=.8]{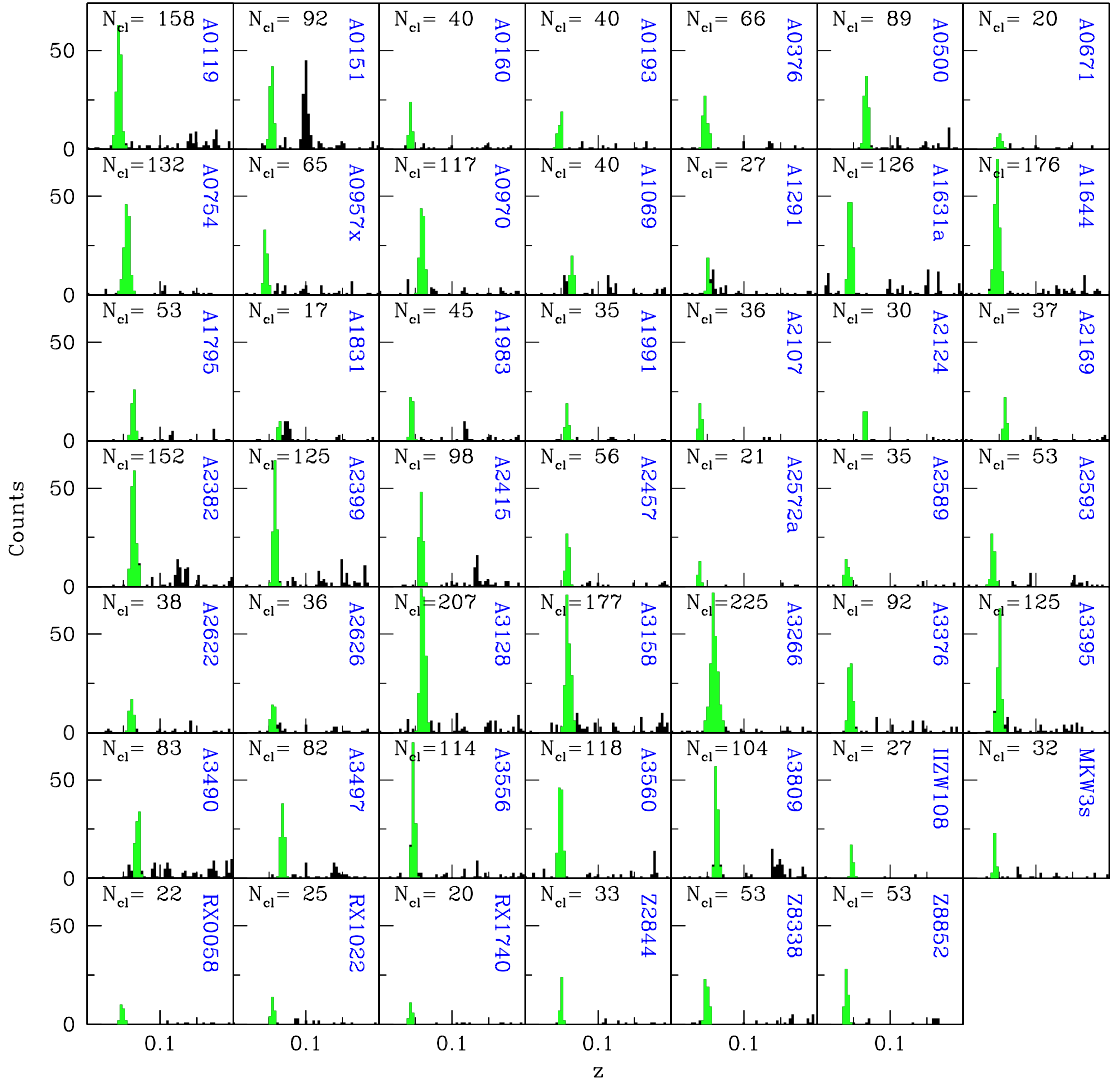}  
  \caption{Histograms of the WINGS-SPE data sample in the wide-range $0\le z \le 0.2$ where fore/background objects are visible. In green are histograms for cluster members defined using a $3\sigma$ cut. $N_{cl}$ indicates the number of cluster galaxies inside $3\sigma$ from the mean cluster redshift.}\label{fig:hist2}
\end{figure*}

In Fig.\ref{fig:hist1}, we show the observed redshift histograms of clusters in the WINGS-SPE sample. In each panel the redshift distributions for the galaxies assigned to the cluster are plotted in green. The bin size in redshift is $0.0015$. A hint of the presence of substructures in some clusters can already be seen here. The same plot but for a larger redshift range ($0\le z \le 0.2$) is given in Fig.\ref{fig:hist2}; the foreground and background galaxies are much more evident here. In some cases a secondary peak indicating a background or a foreground cluster can be seen as well, such as in the case of $A0151$, $A1631a$, and $A2382$. In these cases the bin size is larger and equal to $0.004$.

\begin{figure*}
  \includegraphics[scale=.8]{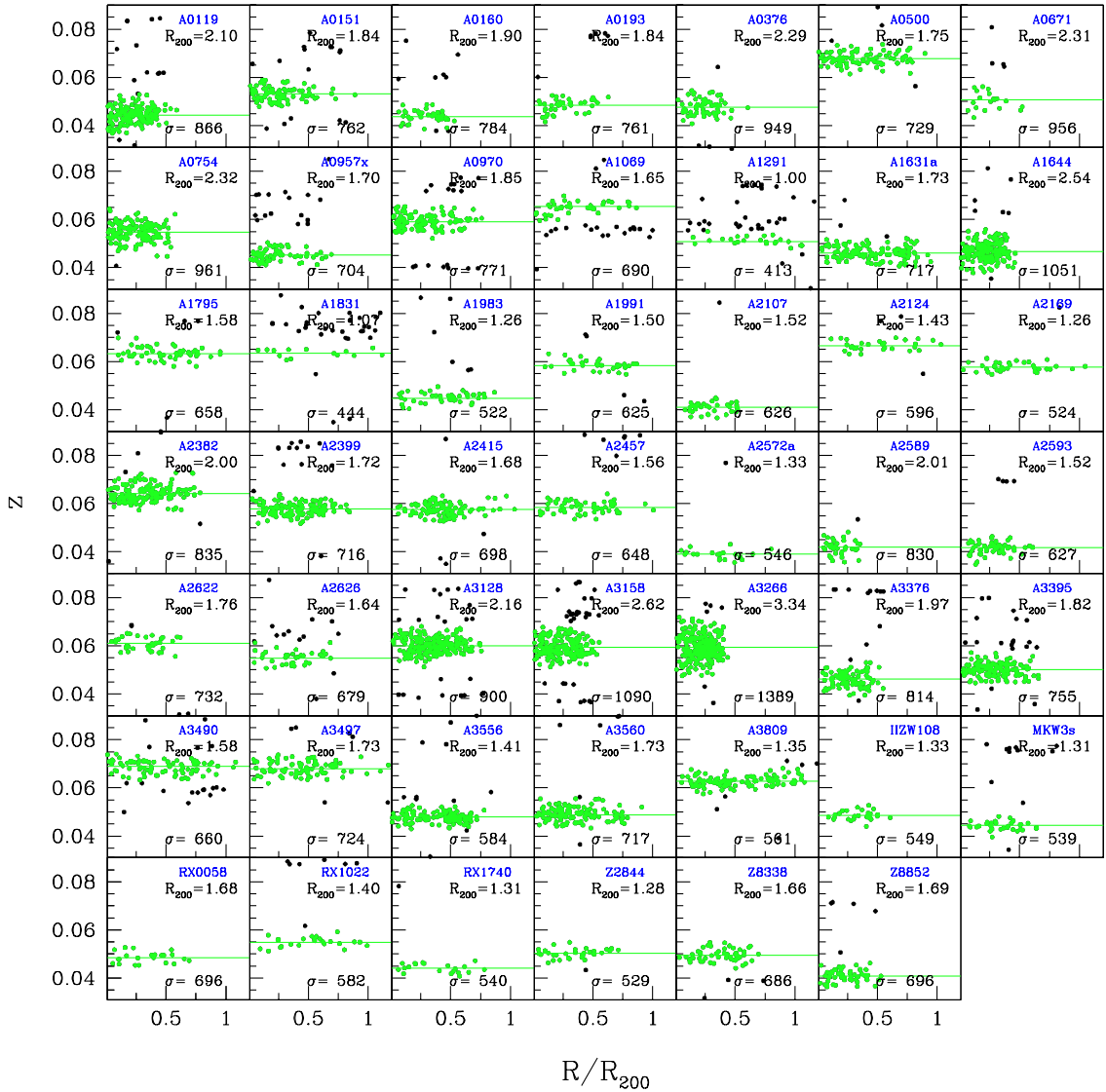}  
  \caption{Velocity diagrams of the WINGS-SPE data sample in the range $0.03\le z \le 0.09$ as a function of the normalized radius ($R/R_{200}$). Green points indicate cluster members, the horizontal line is the mean redshift. $R_{200}$ and $\sigma$ are also indicated in each box. }\label{fig:VRplot}
\end{figure*}

In Fig.\ref{fig:VRplot} we present velocity diagrams for each cluster. Cases where there are additional sub-structures around the main cluster are even more evident in these plots. Green points indicate galaxies assigned as cluster members using the procedure explained above. Also indicated in each plot is the value of $R_{200}$ (in Mpc), the radius inside which the mean density of the cluster is $200$  times the critical density of the universe, and the value of the cluster velocity dispersion, $\sigma_{cl}$ (in km\,s$^{-1}$). $R_{200}$ was computed from $\sigma_{cl}$ as in Poggianti et al. \cite{poggianti06}:
\begin{equation}
\label{pogg}
R_{200}=1.73\frac{\sigma_{cl}}{1000\,kms^{-1}}\,\frac{1}{\sqrt{\Omega_{\Lambda}+\Omega_0(1+z_{cl})^3}}\,h^{-1}\,$Mpc$
\end{equation}
with $\Omega_{\Lambda}=0.7$, $\Omega_0=0.3$, $h=0.7$.\\

With these new data, the mean of the ratio of the number of galaxies we used to derive the cluster velocity dispersion to the number that were used for the literature values is $R=N_{wings}/N_{tab}=2.71$. Hence our measurements are based on almost three times as many member galaxies in comparison to previous measurements. The spectroscopic velocity dispersion for the WINGS-SPE sample are in the range 400 to 1300\,km\,s$^{-1}$, and are generally higher than the velocity dispersions for the SDSS sample at similar redshift (\cite{Miller}).

\subsection{X-ray luminosity-velocity dispersion relation}


\begin{figure}[h!]
\centering\includegraphics[scale=.4]{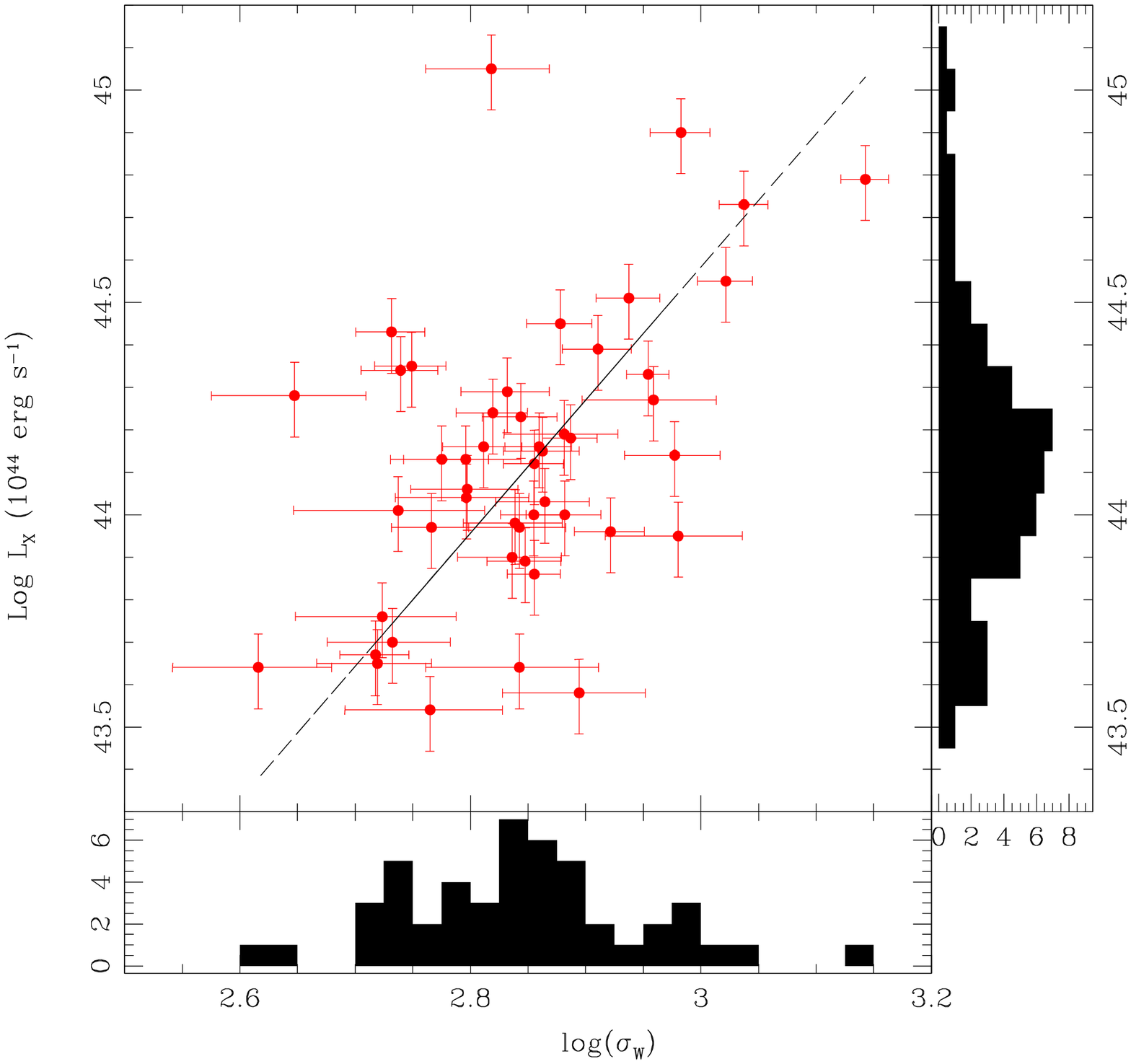}
\caption{X-ray luminosity versus velocity dispersion $\sigma_v$ relation for the 48 clusters in the WINGS sample. The marginal distributions are also shown.}\label{fig:sigma_lx}
\end{figure}

In this section we evaluate the strength and shape of the correlation between X-ray luminosity and velocity dispersion for our sample of clusters. Self-similar models assume that the dominant energy source in the cluster comes from the gravitational collapse, predicting scaling relations of the form: $L_x \propto T^2 \propto \sigma_v^4$. Whereas there seems to be general agreement between different measurements that $L_x\propto T^{\sim 3}$, the measurement  of $L_x-\sigma_v$ has so far given contradictory results. Some authors have found that $L_x \propto \sigma^4$ (although with quite large measurement errors because of rather small data samples), while others find slopes larger than 4 (see e.g. Xue \& Wu 2000, Ortiz-Gil et al. 2004). Some of the differences in the results could arise from the different ways the samples are selected, with a preference for more regular clusters in some of these surveys. 
It has also been suggested that clusters and groups do not follow the same $L_x-\sigma-v$ scaling relation, the latter being flatter than the former (e.g. Mahdavi et al. 2000; Xue and Wu 2000). However, there are  other measurements that contradict that conclusion (e.g. Mulchaey \& Zabludoff 1998; Mahdavi \& Geller 2001). For more distant clusters ($z$ between $\sim$ 0.15 and $\sim$ 0.6) there is some evidence that the slope is also $>4$ (\cite{bor99}; Girardi \& Mezzetti 2001), although only small samples are available at the moment, and more data are needed to reduce the error bars.

In Figure \ref{fig:sigma_lx}, the X-ray luminosities, $L_x$ of the clusters in our WINGS-SPE sample are plotted against their velocity dispersion, $\sigma_w$, with the best fit relation shown. $\sigma_W$ is from our measurements and the X-ray luminosities are taken from \cite{ebeling96} and are in the 0.1-2.4 keV ROSAT RASS bandpass  (see columns (7) and (10) of table \ref{table:prop} respectively).  The marginal distributions of the X-ray luminosities and velocity dispersions are shown in the side panels. Performing an orthogonal fit, the best fit relation is given by: 
\begin{equation}
\log(L_x)=(32.6\pm 1.7) + (4.0\pm 0.3)\times \log(\sigma_v)
\end{equation}
with $L_x$ in units of erg s$^{-1}$ and $\sigma_v$ in km s$^{-1}$.
The observed slope of $4.0\pm0.3$ is in good agreement with the value measured by Mulchaey \& Zabludoff (1998), Mahdavi \& Geller (2001), Girardi \& Mezzetti (2001) and Ortiz-Gil et al. (2004). As for the intercept, the result is compatible with Mahdavi \& Geller (2001), Girardi \& Mezzetti (2001) and Ortiz-Gil et al. (2004) at about one-sigma confidence level.

 \section{Summary}
As part of the WINGS-SPE survey, we have carried out spectroscopic observations of galaxies in 48 clusters using the WHT/AF2 and AAT/2dF facilities. These observations have yielded redshifts for 6,137 galaxies, which have been used to derive precise redshifts and velocity dispersions for our clusters. By combining these data with those already available in the literature, we now have velocity dispersions for all the clusters in the WINGS sample. 

In Table \ref{table:catalogs} we present the complete and final set of redshift data now available for the WINGS galaxy clusters. A total of 3,647 galaxies turn out to be members of our clusters, thereby almost doubling the number of known members in this sample of nearby clusters. 
We have shown that our reduction and measurement procedures result in high quality redshift measurements. A comparison with data available in the literature to both check the accuracy and consistency of our measurements and to increase our overall redshift sample has been done. 
Using an iterative $3\sigma$ clipping scheme, we have derived velocity dispersions for all the 48 WINGS-SPE clusters. The mean of the ratio of the number of galaxies we used to derive the cluster velocity dispersion and the number that were used for the literature values is $R=N_{wings}/N_{tab}=2.71$. This means that our velocity dispersion values are based on almost three times as many member galaxies than previous measurements. 
We found that the X-ray luminosity - velocity dispersion ($L_x$ - $\sigma$) relation for our sample of 48 clusters has $L_x \propto \sigma^4$, although with a large scatter. Finally, we note that the WINGS clusters have a wide range of velocity dispersion values. The implied large range of masses therefore makes the WINGS cluster sample an unprecedented and unique dataset to study the processes affecting cluster galaxy 
evolution as a function of cluster mass. Future papers in this series will exploit this data set to perform a dynamical analysis of the WINGS sample of galaxy clusters and a substructure analysis. 

 \begin{acknowledgements} 
 We wish to thank the anonymous referee for useful comments that have improved this manuscript.
\end{acknowledgements}

\clearpage\onecolumn
\longtab{1}{
\begin{longtable}{lccrccrcllcc}
\caption{The WINGS-SPE cluster sample: global properties} \label{table:prop}\\
\hline \hline
$Cluster$  & $RA$ & $DEC$ & $N_{gal}$ & $z$ & $N_{z}$ & $\sigma_W$ & $\sigma_L$ & Refs &  $log(L_X)$ & $R_{200}$ & $Ap$\\ 
 &$(J2000)$ & $(J2000)$ &  & & & $kms^{-1}$ & $kms^{-1}$ & & $10^{44}erg s^{-1}$ & $Mpc$ & $R_{200}$\\
\hline\\
\endfirsthead
\caption{continued.}\\
\hline\hline
$Cluster$  & $RA$ & $DEC$ & $N_{gal}$ & $z$ & $N_{z}$ & $\sigma_W$  & $\sigma_L$ & Refs &  $log(L_X)$ & $R_{200}$ & $Ap$ \\
 &$(J2000)$ & $(J2000)$ &  & & & $kms^{-1}$ & $kms^{-1}$ & & $10^{44}erg s^{-1}$ & $Mpc$ & $R_{200}$\\
\hline\\
\endhead
\hline
\endfoot
A0119   & 00:56:21.37 & -01:15:46.5 & 248   & 0.0444 & 158   & 866  $^{\pm55}$    & 740       & st99      & 44.51   & 2.10 &  0.6 \\
A0151   & 01:08:52.35 & -15:25:00.9 & 269   & 0.0532 & 92    & 762  $^{\pm57}$    & 669       & st99      & 44.00   & 1.84 &  0.9 \\
A0160   & 01:12:51.40 & +15:30:53.8 & 80    & 0.0438 & 40    & 784  $^{\pm111}$   & 572       & st99      & 43.58   & 1.90 &  0.6 \\
A0193   & 01:25:07.35 & +08:41:35.9 & 62    & 0.0485 & 40    & 761  $^{\pm86}$    & 756       & st99      & 44.19   & 1.84 &  0.6 \\
A0376   & 02:45:48.53 & +36:51:35.5 & 88    & 0.0476 & 66    & 949  $^{\pm90}$    & 519       & st99      & 44.14   & 2.29 &  0.6 \\
A0500   & 04:38:54.97 & -22:05:55.3 & 140   & 0.0678 & 89    & 729  $^{\pm55}$    & $\ldots$  & $\ldots$  & 44.15   & 1.75 &  0.9 \\
A0671   & 08:28:29.28 & +30:25:00.6 & 35    & 0.0507 & 20    & 956  $^{\pm130}$   & 1043      & st99      & 43.95   & 2.31 &  0.5 \\
A0754   & 09:08:50.08 & -09:38:11.8 & 158   & 0.0547 & 132   & 961  $^{\pm57}$    & 931       & st99      & 44.90   & 2.32 &  0.6 \\
A0957x  & 10:13:57.33 & -00:54:54.4 & 128   & 0.0451 & 65    & 704  $^{\pm52}$    & 659       & st99      & 43.89   & 1.70 &  0.7 \\
A0970   & 10:17:34.30 & -10:42:01.5 & 185   & 0.0591 & 117   & 771  $^{\pm42}$    & $\ldots$  & $\ldots$  & 44.18   & 1.85 &  0.8 \\
A1069   & 10:39:54.29 & -08:36:39.8 & 112   & 0.0653 & 40    & 690  $^{\pm68}$    & 1120      & st99      & 43.98   & 1.65 &  0.9 \\
A1291   & 11:32:04.46 & +56:01:26.2 & 85    & 0.0509 & 27    & 413  $^{\pm65}$    & 720       & ag07      & 43.64   & 1.0  &  1.3 \\
A1631a  & 12:52:49.84 & -15:26:17.1 & 227   & 0.0461 & 126   & 717  $^{\pm38}$    & 702       & st99      & 43.86   & 1.73 &  0.9 \\
A1644   & 12:57:14.77 & -17:21:12.6 & 266   & 0.0467 & 176   & 1051 $^{\pm58}$    & 945       & st99      & 44.55   & 2.54 &  0.5 \\
A1795   & 13:49:00.52 & +26:35:06.8 & 91    & 0.0633 & 53    & 658  $^{\pm81}$    & 887       & g96       & 45.05   & 1.58 &  1.0 \\
A1831   & 13:59:10.19 & +27:59:27.9 & 66    & 0.0634 & 17    & 444  $^{\pm68}$    & 316       & st94      & 44.28   & 1.07 &  1.5 \\
A1983   & 14:52:44.00 & +16:44:45.8 & 94    & 0.0447 & 45    & 522  $^{\pm36}$    & 498       & st94      & 43.67   & 1.26 &  0.9 \\
A1991   & 14:54:30.22 & +18:37:51.2 & 50    & 0.0584 & 35    & 625  $^{\pm73}$    & 721       & st99      & 44.13   & 1.50 &  0.9 \\
A2107   & 15:39:47.91 & +21:46:20.6 & 41    & 0.0410 & 36    & 626  $^{\pm83}$    & 625       & g96       & 44.04   & 1.52 &  0.6 \\
A2124   & 15:44:59.33 & +36:03:39.9 & 46    & 0.0666 & 30    & 596  $^{\pm58}$    & 826       & ag07      & 44.13   & 1.43 &  1.0 \\
A2169   & 16:14:06.63 & +49:07:30.6 & 63    & 0.0578 & 37    & 524  $^{\pm60}$    & 521       & ag07      & 43.65   & 1.26 &  1.1 \\
A2382   & 21:52:01.87 & -15:38:53.1 & 247   & 0.0641 & 152   & 835  $^{\pm58}$    & $\ldots$  & $\ldots$  & 43.96   & 2.00 &  0.8 \\
A2399   & 21:57:32.55 & -07:47:40.4 & 242   & 0.0578 & 125   & 716  $^{\pm46}$    & 530       & st99      & 44.00   & 1.72 &  0.9 \\
A2415   & 22:05:25.01 & -05:35:23.1 & 199   & 0.0575 & 98    & 698  $^{\pm52}$    & $\ldots$  & $\ldots$  & 44.23   & 1.68 &  1.1 \\
A2457   & 22:35:45.20 & +01:28:33.3 & 81    & 0.0584 & 56    & 648  $^{\pm51}$    & 316       & st99      & 44.16   & 1.56 &  1.0 \\
A2572a  & 23:18:23.58 & +18:44:24.7 & 26    & 0.0390 & 21    & 546  $^{\pm103}$   & 676       & st99      & 44.01   & 1.33 &  0.8 \\
A2589   & 23:24:00.52 & +16:49:29.0 & 47    & 0.0419 & 35    & 830  $^{\pm98}$   & 819       & st99      & 44.27    & 2.01 &  0.4 \\
A2593   & 23:24:31.01 & +14:38:29.3 & 86    & 0.0417 & 53    & 627  $^{\pm67}$    & 763       & st99      & 44.06   & 1.52 &  0.7 \\
A2622   & 23:34:53.81 & +27:25:35.5 & 71    & 0.0610 & 38    & 732  $^{\pm68}$    & 942       & st99      & 44.03   & 1.76 &  0.6 \\
A2626   & 23:36:31.00 & +21:09:36.3 & 70    & 0.0548 & 36    & 679  $^{\pm60}$    & 696       & st99      & 44.29   & 1.64 &  0.7 \\
A3128   & 03:30:34.63 & -52:33:12.2 & 297   & 0.0600 & 207   & 900  $^{\pm38}$    & 802       & st99      & 44.33   & 2.16 &  0.8 \\
A3158   & 03:42:39.64 & -53:37:50.1 & 278   & 0.0593 & 177   & 1090  $^{\pm53}$   & 976       & st99      & 44.73   & 2.62 &  0.6 \\
A3266   & 04:31:11.92 & -61:24:22.7 & 264   & 0.0593 & 225   & 1389  $^{\pm66}$   & 1085      & st99      & 44.79   & 3.34 &  0.4 \\
A3376   & 06:00:43.57 & -40:02:59.5 & 144   & 0.0461 & 92    & 814  $^{\pm56}$    & 641       & st99      & 44.39   & 1.97 &  0.6 \\
A3395   & 06:27:31.09 & -54:23:57.8 & 191   & 0.0500 & 125   & 755  $^{\pm49}$    & 1090      & st99      & 44.45   & 1.82 &  0.7 \\
A3490   & 11:45:18.58 & -34:26:40.0 & 218   & 0.0688 & 83    & 660  $^{\pm47}$    & $\ldots$  & $\ldots$  & 44.24   & 1.58 &  1.1 \\
A3497   & 12:00:03.53 & -31:23:42.4 & 165   & 0.0680 & 82    & 724  $^{\pm48}$    & $\ldots$  & $\ldots$  & 44.16   & 1.73 &  1.3 \\
A3556   & 13:24:06.23 & -31:39:37.8 & 175   & 0.0479 & 114   & 584  $^{\pm45}$    & 643       & st99      & 43.97   & 1.41 &  0.8 \\
A3560   & 13:31:50.50 & -33:13:25.4 & 191   & 0.0489 & 118   & 717  $^{\pm43}$    & 1123      & st99      & 44.12   & 1.73 &  0.9 \\
A3809   & 21:46:51.76 & -43:52:54.7 & 195   & 0.0627 & 104   & 561  $^{\pm40}$    & 499       & st99      & 44.35   & 1.35 &  1.1 \\
IIZW108 & 21:13:56.00 & +02:33:56.0 & 32    & 0.0483 & 27    & 549  $^{\pm42}$    & $\ldots$  & $\ldots$  & 44.34   & 1.33 &  0.6 \\
MKW3s   & 15:21:50.00 & +07:42:32.0 & 66    & 0.0444 & 32    & 539  $^{\pm37}$    & 612       & g96       & 44.43   & 1.31 &  0.7 \\
RX0058  & 00:58:22.30 & +26:52:03.7 & 31    & 0.0484 & 22    & 696  $^{\pm119}$   & $\ldots$  & $\ldots$  & 43.64   & 1.68 &  0.7 \\
RX1022  & 10:22:07.30 & +38:30:55.0 & 44    & 0.0548 & 25    & 582  $^{\pm91}$    & 591       & ag07      & 43.54   & 1.40 &  1.0 \\
RX1740  & 17:40:32.30 & +35:38:57.0 & 32    & 0.0441 & 20    & 540  $^{\pm66}$    & $\ldots$  & $\ldots$  & 43.70   & 1.31 &  0.8 \\
Z2844   & 10:02:37.09 & +32:41:16.7 & 54    & 0.0503 & 33    & 529  $^{\pm84}$    & $\ldots$  & $\ldots$  & 43.76   & 1.28 &  0.7 \\
Z8338   & 18:11:26.75 & +49:49:47.4 & 86    & 0.0494 & 53    & 686  $^{\pm71}$    & $\ldots$  & $\ldots$  & 43.90   & 1.66 &  0.7 \\
Z8852   & 23:10:31.85 & +07:34:17.7 & 71    & 0.0408 & 53    & 696  $^{\pm67}$    & $\ldots$  & $\ldots$  & 43.97   & 1.69 &  0.5 \\
\hline\hline
\end{longtable}
{\small\noindent Refs:
 (st94)  Struble and Ftaclas 1994;
 (g96) Girardi et al. 1996;
 (maz96) Mazure et al. 1996;
 (wu98)  Wu et al. 1998;
 (st99)  Struble and Rood 1999;
 (FPS04) Smith et al. 2004;
 (ag07)  Aguerri et al. 2007\\
Columns: (1) cluster name, (2-3) coordinates of the image field center at
epoch $2000$ (right ascension (2) in hours, minutes and seconds and
declination (3) in degrees, arcminutes, arcseconds ), (4) number of
redshift determinations which is equal to the number of entries in the
spectroscopic catalog for that cluster, (5) cluster mean redshift, (6)
number of member galaxies (used to compute mean redshift and velocity
dispersion as explained in the text), (7) cluster velocity dispersion computed from WINGS data, (8) cluster velocity dispersion from literature, (9) reference for the literature velocity dispersion, (10) logarithm of the X-ray luminosity in the $0.1-2.4\,keV$ ROSAT RASS bandpass (from \cite{ebeling96}), (11) virial radius in $Mpc$, (12) aperture in units of $R_{200}$.}
} 
 
\longtab{2}{
\begin{longtable}{lccrccrcll}
\caption{Global properties of the remaining $29$ clusters of the WINGS sample} \label{table:prop2}\\
\hline \hline
$Cluster$  & $RA$ & $DEC$ & $z$ & $N_{z}$ & $\sigma_L$ & Refs & $log(L_X)$\\ 
 &$(J2000)$ & $(J2000)$ & & & $kms^{-1}$ & & $10^{44}erg s^{-1}$ \\
\hline\\
\endfirsthead
\caption{continued.}\\
\hline\hline
$Cluster$  & $RA$ & $DEC$ & $z$ & $N_{z}$ & $\sigma_L$ & Refs & $log(L_X)$\\
 &$(J2000)$ & $(J2000)$ &  & & $kms^{-1}$ & &$10^{44}erg s^{-1}$ \\
\hline\\
\endhead
\hline
\endfoot
A0085   & 00:41:37.81  & -09:20:33.2    & 0.0521    &  273 & 979       & ag07 & 44.92 \\
A0133   & 01:02:38.97  & -21:57:15.4    & 0.0603    &  7   & 623       & st99 & 44.55 \\
A0147   & 01:08:10.44  & +02:09:59.9    & 0.0447    &  11  & 387       & st99 & 43.73 \\
A0168   & 01:15:09.80  & +00:14:50.6    & 0.0448    &  106 & 578       & ag07 & 44.04 \\
A0311   & 02:09:10.34  & +19:43:10.4    & 0.0657    &  1 & \ldots      & wu98 & 43.91 \\
A0548b  & 05:47:01.74  & -25:36:58.8    & 0.0441    &  323 & 842       & st99 & 43.48 \\
A0602   & 07:53:19.02  & +29:21:10.5    & 0.0621    &  78  & 834       & ag07 & 44.05 \\
A0780   & 09:18:30.36  & -12:15:40.1    & 0.0565    &  34  & $\ldots$  & FPS04 & 44.82 \\
A1668   & 13:03:51.41  & +19:15:55.1    & 0.0634    &  15  & 654       & st99 & 44.20 \\
A1736   & 13:26:52.16  & -27:06:33.5    & 0.0461    &  109 & 918       & st99 & 44.37 \\
A2149   & 16:01:38.10  & +53:52:42.9    & 0.0675    &  20  & 459       & ag07 & 43.92 \\
A2256   & 17:03:43.53  & +78:43:02.6    & 0.0581    &  116 & 1376      & st99 & 44.85 \\
A2271   & 17:17:17.53  & +78:01:00.0    & 0.0584    &  10  & 460       & st99 & 43.80 \\
A2657   & 23:44:51.00  & +09:08:39.6    & 0.0400    &  31  & 829       & st99 & 44.20 \\
A2665   & 23:50:45.44  & +06:06:41.2    & 0.0562    &   2  & $\ldots$  & st99 & 44.28 \\
A2717   & 00:02:59.40  & -36:02:05.5    & 0.0498    &  56  & 512       & st99 & 44.00 \\
A2734   & 00:11:20.13  & -28:52:18.5    & 0.0624    &  80  & 628       & st99 & 44.41 \\
A3164   & 03:45:49.70  & -57:02:43.9    & 0.0611    &   3  & 991       & st99 & 44.17 \\
A3528a  & 12:54:31.28  & -29:22:21.6    & 0.0535    &  28  & 969       & maz96 & 44.12 \\
A3528b  & 12:54:08.64  & -28:50:32.5    & 0.0535    &  6   & $\ldots$  & FPS04 & 44.30 \\
A3530   & 12:55:36.88  & -30:21:14.4    & 0.0544    &  $\ldots$ & 391  & wu98 & 43.94 \\
A3532   & 12:57:19.17  & -30:22:12.7    & 0.0555    &  44  & 742       & st99 & 44.45 \\
A3558   & 13:27:54.77  & -31:29:31.8    & 0.0477    &  341 & 977       & st99 & 44.80 \\
A3562   & 13:33:31.81  & -31:40:22.5    & 0.0502    &  114 & 1048      & st99 & 44.52 \\
A3667   & 20:12:30.09  & -56:48:59.5    & 0.0530    &  162 & 1059      & st99 & 44.94 \\
A3716   & 20:51:16.71  & -52:41:43.5    & 0.0448    &  111 & 733       & st99 & 44.00 \\
A3880   & 22:27:49.53  & -30:34:40.0    & 0.0570    &  22  & 855       & st99 & 44.27 \\
A4059   & 23:56:40.70  & -34:40:17.7    & 0.0480    &  45  & 628       & FPS04 & 44.49 \\
Z1261   & 07:16:42.60  & +53:24:28.0    & 0.0644    &  $\ldots$ & $\ldots$  & wu98  &43.91 \\
\hline\hline
\end{longtable}
{\small\noindent Refs: 
 (st94)  Struble and Ftaclas 1994;  
 (g96) Girardi et al. 1996; 
 (maz96) Mazure et al. 1996; 
 (wu98)  Wu et al. 1998; 
 (st99)  Struble and Rood 1999; 
 (FPS04) Smith et al. 2004; 
 (ag07)  Aguerri et al. 2007\\
Columns: (1) cluster name, (2-3) coordinates of the image field center at
epoch $2000$ (right ascension (2) in hours, minutes and seconds and
declination (3) in degrees, arcminutes, arcseconds ), (4) cluster mean redshift, (5) number of member galaxies, (6) cluster velocity dispersion (7) reference for the data reported in columns (4-6), (8) logarithm of the X-ray luminosity in the $0.1-2.4\,keV$ ROSAT RASS bandpass (from \cite{ebeling96})}
} 
\end{document}